# Deep Learning and Knowledge-Based Methods for Computer-Aided Molecular Design — Toward a Unified Approach: State-of-the-Art and Future Directions

Abdulelah S. Alshehri[a,b], Rafiqul Gani[c], Fengqi You[a*]

[a] Robert Frederick Smith School of Chemical and Biomolecular Engineering, Cornell University, Ithaca, NY 14853, USA

[b] Department of Chemical Engineering, College of Engineering, King Saud University, P.O. Box 800, Riyadh 11421, Saudi Arabia

[c] PSE for SPEED Company, Skyttemosen 6, DK-3450 Allerød, Denmark

July 5, 2020

## Abstract

The optimal design of compounds through manipulating properties at the molecular level is often the key to considerable scientific advances and improved process systems performance. This paper highlights key trends, challenges, and opportunities underpinning the Computer-Aided Molecular Design (CAMD) problems. A brief review of knowledge-driven property estimation methods and solution techniques, as well as corresponding CAMD tools and applications, are first presented. In view of the computational challenges plaguing knowledge-based methods and techniques, we survey the current state-of-the-art applications of deep learning to molecular design as a fertile approach towards overcoming computational limitations and navigating uncharted territories of the chemical space. The main focus of the survey is given to deep generative modeling of molecules under various deep learning architectures and different molecular representations. Further, the importance of benchmarking and empirical rigor in building deep learning models is spotlighted. The review article also presents a detailed discussion of the current perspectives and challenges of knowledge-based and data-driven CAMD and identifies key areas for future research directions. Special emphasis is on the fertile avenue of hybrid modeling paradigm, in which deep learning approaches are exploited while leveraging the accumulated wealth of knowledge-driven CAMD methods and tools.

*Corresponding author. Phone: (607) 255-1162; Fax: (607) 255-9166; E-mail: fengqi.you@cornell.edu

## 1. Introduction

Materials have been a pivotal part of the modern economy and a prospective solution route to solving the world's most pressing scientific problems that span disciplines and impact communities [1], [2]. Designing molecules to improve the functionality and efficiency of products and processes can bring tangible environmental, technological, and economic benefits in energy harvesting and storage, medical diagnostics and therapy, and carbon capture and utilization, to mention but a few. The design process of chemicals involves deriving new versions from existing molecules or creating original ones from novel molecules [1]. Yet, molecular design has been a daunting trial-and-error experiment-based or/and heuristic rule-based process constrained by resources and limited to a small class of known molecular structures [3], [4]. In spite of century-long efforts in chemical synthesis and the large set of synthesized molecules (~$10^7$), the so-called chemical space is still an unexplored galaxy with an estimated number of small organic molecules populating the space of more than $10^{60}$ [5], [6]. Revolutionary advances are likely to emerge from novel molecules in the uncharted territories of the chemical space.

Efforts to systematically characterize and link properties to molecules root back to the 1940s with the development of an early group contribution (GC) method, predicting heat and free energies of organic compounds [7]. A recent GC-based model was successful in reaching chemical accuracy for the enthalpy of formation of organic compounds. To some extent, advancements in property prediction methods have been driven by the need to build a framework for its reverse problem. This ultimate purpose was best emphasized by the organic photochemist, George S. Hammond, in his speech upon receiving the Norris Award in Physical Organic Chemistry in 1968, "The most fundamental and lasting objective of synthesis is not production of new compounds, but production of properties" [8]. A few decades later, the availability of powerful computational resources in conjunction with the growing demand for chemical products gave rise to the field of computer-aided molecular design (CAMD), which combines property estimation methods with mathematical optimization methods to find promising molecular structures that best meet target properties [9]. In this context, GC-based methods are the most used class of property estimation methods in CAMD due to their easy incorporation within mathematical models and high qualitatively correct estimates [10].

Identified as a key challenge in the new millennium, the importance of CAMD stems from the value and potential unlocked by connecting the molecular-level design with properties at the macroscopic level [11]. Propelled partly by the exponential growth in the computing power and partly by advances in mathematical optimization algorithms and property estimation models, CAMD has attracted growing research activities and enjoyed enormous progress across multiple scales — ranging from single-molecule design to mixture design and integrated product and process design. Yet, knowledge-based methods either suffer from the inadequate number of available property models or require significant computational resources to explore the staggeringly vast discrete design space. As such, severe limitations on the quality of solutions are imposed by the exponential growth of complexity with the size of molecules. Hence, the growing market demand for specialty chemicals and the global environmental concerns highlights the urgent need for smarter, more efficient, and more effective methods to navigate the chemical space in order to accelerate innovation and discovery in molecular design.

Machine learning has emerged to outperform many conventional algorithms and artificial intelligence techniques across various fields, such as computer vision [12], [13], speech recognition [14], [15], and natural language processing [16], [17]. In particular, the subfield of deep learning has shown a remarkable ability to discover intricate structures in high-dimensional data and transform complex representations into higher abstract levels [18]. For the problem of property prediction, deep learning has not only beaten other machine learning methods for building an accurate QSPR model [19], but also has recently closed the gap between QSPR predictive models and the quantum chemistry-based Density Functional Theory (DFT) on the formation energy of materials [20]. As mathematical optimization plays a central role in CAMD, deep learning has enabled unforeseen transformative advances on the theoretical and practical aspects of optimization, as in the linearization of mixed-integer quadratic programming problems [21], optimization under uncertainty [22], human-level control [23], and battery charging protocols [24]. Further, deep generative models have recently made large strides in their ability to model molecular distributions and synthesize realistic molecules from learned distributions, as evidenced by numerous successful applications in molecular design [25], [26]. The promise of integrating deep learning in CAMD is the better performance of computational models that offers a more extensive and less demanding landscape for optimization and characterization, leading to novel technologies and reductions in computational demand [27]. Several sources in the literature have pointed out potential hybrid routes for harnessing the promise of deep learning, while leveraging the accumulated wealth of scientific knowledge by developing theory-guided neural networks in quantum chemistry [28], fluid mechanics [29], and system dynamics [30].

A number of excellent reviews of the knowledge-based CAMD paradigm are available in the literature [4], [9], [31], [32]. Instead of merely focusing on knowledge-based methods and techniques, this review lays out a comparative description of both the knowledge-based CAMD the emerging deep learning approaches, identifies challenges and gaps in the approaches, and explores opportunities and future directions of these approaches and their combination. In the following section, we describe and discuss the key principles of the knowledge-based CAMD paradigm, including property estimation methods and solution techniques for organic molecules, along with their tools and applications. Section 3 surveys the current state-of-the-art in the nascent area of deep learning for molecular design, covering three main elements: molecular representations, major deep generative architectures, and benchmarking and evaluation metrics. The survey starts by assessing the relative merits of different molecular representations, which in turn define the information and structures that can be exploited in learning-based model development. It is followed by a discussion detailing the mathematical structure, respective strengths, and reported results of several deep learning architectures. Further, special emphasis is placed on benchmarks for evaluating the generation of molecules and the optimization of their properties in learning-based models. In section 4, we offer a thorough discussion of the perspectives and challenges of the knowledge-based and data-driven CAMD paradigms to identify important future research directions for the rapidly evolving field. We also highlight the potential of a hybrid modeling paradigm combining the strengths of knowledge-driven CAMD methods and powerful deep learning techniques for more effective, accurate reliable, and efficient molecular design. The stages of the classical knowledge-based CAMD and recently developed data-driven frameworks are illustrated in Figure 1.

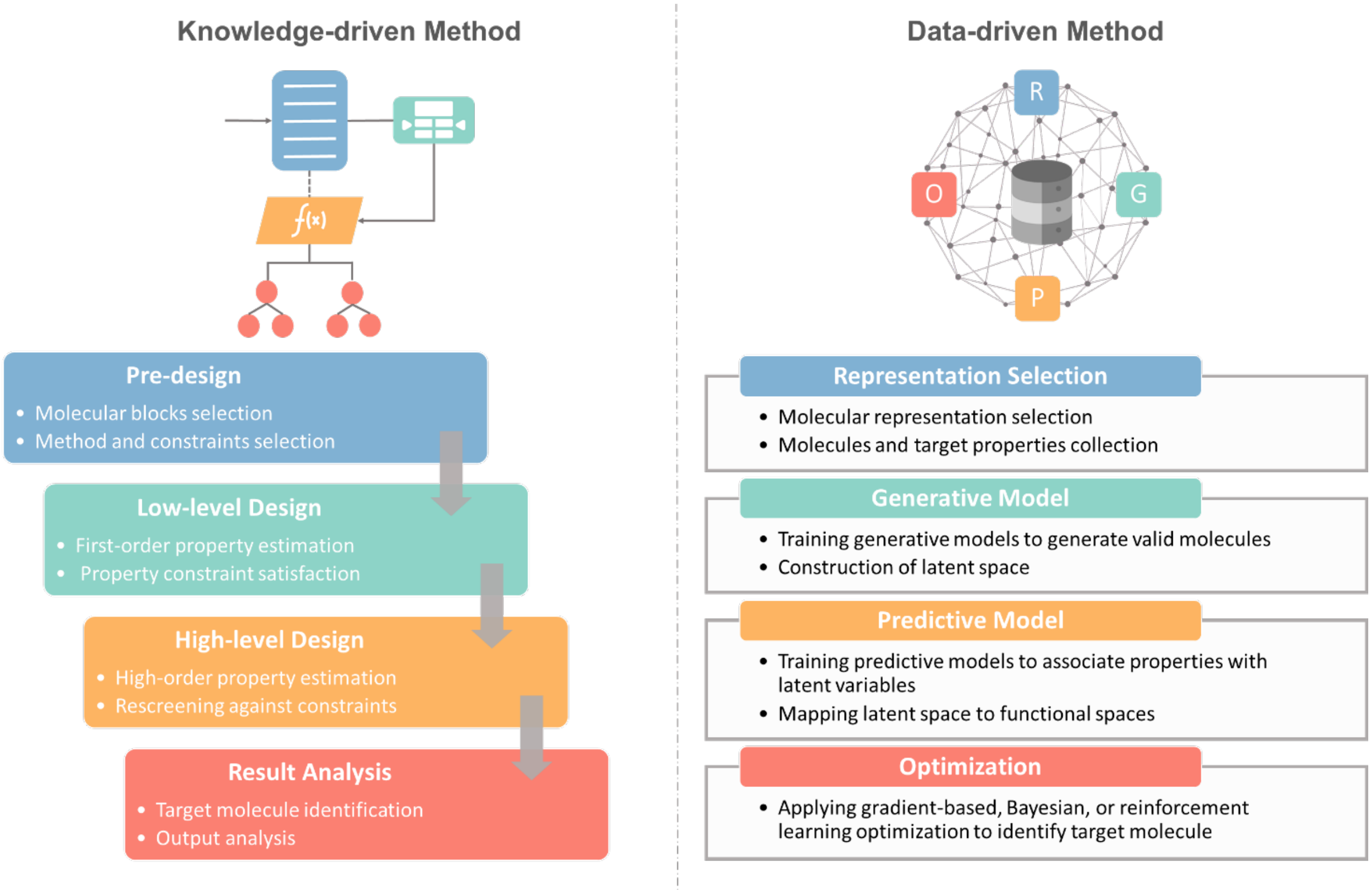

**Figure 1.** Diagrammatic comparison of computer-aided molecular design methods.

## 2. Knowledge-based Computer-Aided Molecular Design

A generic solution framework for CAMD involves two components for molecular modeling: (1) a model for the estimation of molecular properties and the thermodynamics of their mixtures, and (2) computational optimization algorithms for guiding the search in the large, discrete chemical space. In this section, we summarize several foundational aspects to the problem of CAMD: property prediction methods for pure organic components and mixtures, mathematical formulations of CAMD classes, solution methods and techniques, software tools, and common applications. This section also offers comparative descriptions of the knowledge-based CAMD framework components.

### 2.1. Property Estimation Methods

Finding a solution to a CAMD problem is dependent on the ability to efficiently and accurately estimate target properties [33]. The estimation involves the construction of pure compound models and functional and thermodynamic models for mixtures. Despite the staggering number of mechanistic, empirical, and semi-empirical property estimation methods, most of them are not applicable in CAMD. Semi-empirical models have been the most popular class of methods due to their lower computational cost and the representation of molecules that can be easily encoded within optimization models [4], [34]. However, such models enforce further constraints on the formulation of the CAMD problem in the form of application ranges, availability of parameters, and inherent uncertainty within the estimation models. The generation of practical CAMD solutions typically follows a hierarchical ranking process. In the ranking process, the computational effort is adjusted as the search space is narrowed in subsequent abstraction levels, shifting property prediction from simple GC-based methods to complex computations and experimental measurements [3].

At the most essential level, the CAMD and property prediction problems require molecules to be decomposed into their basic chemical building units of atoms, bonds, rings, functional groups, atomic types, charges, and so on. The various representations of building units give rise to different QSPR methods. It should be noted that there is a distinction between the GC family of methods and QSPRs although GC-based can be classified as a special case of QSPRs [10], [35]. In this context, QSPR with descriptors can characterize property very well whereas GC methods offer a parallel performance with fewer parameters [10]. In the next subsections, widely-applied QSPR methods for pure components and mixture property models are summarized and comparatively evaluated. We note that the following property prediction discussion revolves around organic molecules and systems despite recent extensions of these methods to other chemicals as ionic liquids [36] and electrolytes [37].

### 2.1.1. Pure Component Quantitative Structure-Property Relationships

Group contribution (GC) methods, topological indices (TI), and signature descriptors (SD) are the most popular semi-empirical models for pure component property prediction in CAMD. Particularly, most methods in CAMD applications are based on the GC or additivity approach. First proposed in 1949, this approach is probably the earliest method for estimating several properties of compounds [7]. Additivity or contribution methods express properties of molecules as functions of the number of occurrences of molecular fragments called functional groups [10], [38]. The accuracies of earlier GC models have been significantly improved due to the development of more elaborate models that include polyfunctional and structural groups and interaction terms. GC-based models have proven capable of achieving high levels of accuracy, estimating the enthalpy of formation for or a broad range of organic molecules within chemical accuracy [39]. A prevalent GC method in CAMD applications is the GC+ method, which performs the estimation at three distinct levels: a first-order using contributions from simple groups, *F*; a second-order for polyfunctional compounds and identification of isomers, *S*; and a third-order for overreaching structural features, such as fused rings, *T* [33], [40]. Improved models and additional features of the GC+ method have been proposed, such as revised parameters with uncertainty estimates [41] and 22 environment-related properties [42]. For a vector of the number of occurrences of each group $n_g$ and a vector of regression coefficients $c_g$, the GC+ model has the form of the following equation:

$$P = f\left(\sum_{g\epsilon F} c_g n_g + \sum_{g\epsilon S} c_g n_g + \sum_{g\epsilon T} c_g n_g\right) \quad (1)$$

The GC+ model includes 182 first-order, 122 second-order, and 66 third-order groups. Past GC-based models have suffered from degradation in the accuracy of large multifunctional molecules [10]. However, recent models have addressed this limitation, reporting acceptable performance for complex molecules like amino acids and lipids [43], [44]. We point the interested reader to a recent review for more detailed coverage of the current limitations and opportunities in GC methods [10].

TI and SD QSPR methods are based on the chemical graph theory, where the atoms and bonds of a chemical structure are represented as nodes and edges in a graph [45]. TIs are computed using functions that capture several properties in the chemical graph as connectivity and atomic types [46]. Signatures in SDs are systematic codification procedures over an alphabet of atomic types, characterizing the neighborhood of the atoms present in a molecule [47]. In these classes of QSPRs, instead of pairing groups with regression coefficients as in GC, indices and descriptors are used to relate properties to different molecular graph features. TI takes several forms and functions, such as the sum of the shortest distances between all pairs of nodes, known as the Wiener index [48], and the sum of bond contributions in the Randić index [49]. Similar to TIs, SDs translate molecular graphs into molecular descriptors, but SDs preserve connectivity and structural information through node coloring that differentiates between different atoms and different types of same atoms [4]. It should be noted that both the TI and SD methods have limited applications in CAMD [4].

When comparing these popular approaches for CAMD applications, it is useful to focus on their predictive and discriminative powers, together with their applicability within optimization models. As we discuss the generalizability of predictions and the robustness of regression variables, it is essential to highlight that chemical graph methods are more reflective of the molecular structure [45]. This is owed to their expression as functions of the entire chemical graph as opposed to independently contributing groups as in the GC class of methods. Nevertheless, this highly descriptive nature and the lack of established models for wider classes of molecules of TIs and SDs significantly degrade their predictive power, leading to overfitting issues and several limitations on the design space. Despite this downside on the predictive power, TIs and SD are far more discriminatively powerful than GC as demonstrated by their ability to discriminate between similar chemical structures including stereoisomers. Still, for general CAMD problems, chemical graph-based QSPR methods are more problematic to incorporate into optimization models, rendering some problems intractable as they require more binary and discrete variables [4]. It is worth noting that the size and complexity issues have been alleviated by decomposition-based methods, which are covered in subsection 2.3.4.. In these methods, the pre-analysis stage of the framework substantially reduces the search space by removing redundant binary variables [34].

### 2.1.2. Mixture Property Models

For the problem of mixture property prediction, properties can be classified into functional and equilibrium-based. In the case of functional properties, the estimation is based on pure component QSPR methods and a mixing rule for a given set of molecules and their compositions with a predefined phase identity. In the second class, equilibrium-based properties are predicted using calculation algorithms for vapor-liquid equilibrium (VLE), liquid-liquid equilibrium (LLE), and solid-liquid equilibrium (SLE). Generating estimates for mixture properties requires the integration of pure component and equilibrium-based property models to predict properties and calculate phase behaviors [9], [50]. Mixture thermodynamic models have been well-studied across different disciplines with numerous applied models in CAMD including UNIFAC [51], SAFT [52], and COSMO-based methods [40], [53]. In the problem of mixture design, the use of COMSO-based methods is advantageous over other methods as they do not require binary interaction parameters, which pose a challenge to the search algorithm for design optimization. Moreover, the collection of COSMO methods provides an accuracy level comparable to the quantum chemistry-based method, DFT [4], [53].

## 2.2. Mathematical Optimization Formulation for CAMD

Several alternative optimization formulations of the CAMD problem have been developed in the literature to address the problem at different scales and design objectives. The levels of the problem and their applications fall broadly into three classes: single-molecule design, mixture/blend design, and integrated process and product design. The last design class encloses both or either of the other two classes into a process/product problem with an explicit relationship between the molecule/mixture and the process/product. A generic mathematical formulation of the integrated process/product design problem is formulated as the following mixed-integer nonlinear programming (MINLP) model [31]:

$$\max_{x,z,y,w,v} C^T z + f(x) \tag{2}$$

s.t.

$$P_r = f(x,z,w,v,\theta) \tag{3}$$

$$L_1 \le \theta_1(y,\phi) \le U_1 \tag{4}$$

$$L_2 \le \theta_2(y,\phi,\theta) \le U_2 \tag{5}$$

$$L_3 \le \theta_3(y,\phi,\theta,x) \le U_3 \tag{6}$$

$$L_4 \le \theta_4(y,\phi,\theta,x,z) \le U_4 \tag{7}$$

$$S_L \le S(y,\eta) \le S_U \tag{8}$$

$$Bx + C^T z \le D \tag{9}$$

Where **x** is a vector of continuous variables (e.g., operations conditions, flow rate, etc.), **z** is a vector of measured-controlled variables (e.g., temperature, pressure, etc.), **y** is a vector of binary variables (e.g., descriptor identity, molecule identity, etc.), **w** and **v** are vectors of input variables and manipulated-design variables, respectively. Eq. (3) expresses the process model equations as functions of the model variables and the property functions, $\theta$. Eqs. (4)-(7) provide upper and lower bounds on the different classes of pure component and mixture properties with $\phi$ as a structural parameters vector. The last two constraints ensure the molecular and flowsheet feasibility, respectively, given the constant vectors $B$ and $D$ [50].

The above model represents an integrated process and product design problem, but many variations of CAMD problems can be derived from the model. For instance, the subset of Eqs. (4)-(8) gives the molecular feasibility problem, where molecules that satisfy target properties are generated and tested for structural validity. The addition of the cost function to the feasibility problem turns it into the single-molecule design problem that could determine the optimal molecular structure [31].

## 2.3. CAMD Solution Methods

### 2.3.1. Enumeration Methods

Early efforts in the search for molecules and mixtures for specific applications have been carried out by enumeration methods. Solving the CAMD problem using this class of approaches follows a combinatorial approach. That is generating chemically feasible molecules and estimating their target properties, followed by screening the properties against specified property constraints and ranking the molecules using an evaluation metric [50]. Given the low computational burden of evaluating properties from QSPRs, this sequential "generate-and-test" approach is particularly efficient for selecting an optimal molecule over a small pool of molecules in the chemical space. Yet, the major obstacle of applying this class of solution methods lies in limiting the design space to a practical size for larger problems. The combinatorial explosion problem that arises in large design spaces can be circumvented by controlling the generation and testing steps through screening. A number of algorithms have been developed to incorporate knowledge-based evaluation and sequential interval analysis for efficiently reducing the number of candidate molecules [54], [55].

### 2.3.2. Mathematical programming

In addition to enumeration methods, the field of CAMD has benefited from the rapid advances in mathematical optimization algorithms in identifying solutions to problems with nonconvexities and nonlinearities in the molecular system model. This class of methods guarantee optimality for convex problems and serves as a theoretical basis for heuristics and decomposition methods [4]. Before presenting a few relevant optimization frameworks for CAMD problems, we note the challenges faced in the development of the CAMD algorithmic framework. Sharing the same challenge as the enumeration methods, the binary variables associated with descriptors and structural interactions give rise to a combinatorial explosion. Even for the 182 first-order descriptors in Eq. (1), the number of possible group combinations for a maximum molecule size of 15 descriptors grows to the order of $10^{23}$. Moreover, current representations of molecular structures in CAMD are plagued by redundancy, where each molecule is not given by a single unique representation [56]. These challenges are further compounded by the inherent complexity of synthesis and the cost of production. To address these barriers, many researchers have recognized the need for strategy-oriented techniques, ranging from exploiting the problem structure to altering its formulation. Early efforts involve exploiting the problem structure to implement exact optimization algorithms as the branch-and-reduce algorithm [57], [58]. A noteworthy exploitation involved branching on molecular descriptors and property values to efficiently compute all feasible solutions using a single branching tree in conjunction with a feasibility pre-solver [59]. Other attempts rely on the application of reformulation techniques and the development of multi-stage frameworks to sidestep nonlinearity and nonconvexity while guarding against the exclusion of possible optimal solutions from the design space [60], [61]. Developments in this direction include the extension of a decomposition method [56] through the simultaneous consideration of the first-order and second-order functional groups, illustrating the conservation of the original feasible space [61]. Other improvements came about through the introduction of novel constraints [59], [62], [63], and applying reformulation methods using linearization techniques [56], [64], [65] and convexified terms [60], [66].

### 2.3.3. Derivative-free Optimization

When the chemical design space is too large to be handled by exact optimization algorithms, an intuitive alternative to knowledge-based frameworks is a high-level search strategy that guides the search process. Under the absence of an algebraic form of the optimization problem, this class of black-box methods searches the design space either stochastically using adaptive operations, or deterministically by following a specific set of operations. Stochastic methods or metaheuristics apply high-level selection strategies to sample the chemical space, evaluating the objective function for each sample point, and exploiting the accumulated search experience to identify regions in the chemical space with high-quality molecules [9]. Alternatively, given a specific starting point, deterministic methods evaluate molecules with fixed procedures and arrive at the same final molecule [67]. Within these fixed procedures, different variants of local and global search methods are used, such as trust-region methods [68] and SNOBFIT [69]. The two subclasses of Derivative-Free Optimization (DFO) methods can be applied both directly without an underlying model, and indirectly in connection with a surrogate model that offers derivative approximations [70].

In CAMD, many DFO approaches have been applied to optimize the generation of molecular structures. For the stochastic subclass of methods, the natural selection-based genetic algorithms are by far the most commonly used metaheuristic in CAMD, with applications that span from encoding molecular structures to optimizing discrete variables in integrated product and process design problems [71]–[74]. Moreover, Tabu search algorithms have proven highly effective in molecular design problems as metal catalysts, ionic liquids, and polymers [75]–[78]. On the other side, deterministic methods have only been adopted recently and to a limited extent [67], [79]. A study comparing 27 DFO algorithms on the problem of mixture design found global and surrogate-based methods to perform better than other solution strategies [67]. However, it should be noted that the nature of DFO algorithms causes the quality of solutions to be highly dependent on the spaces and time limits, and hyperparameters that govern the search. Further, developed models tend to be only applicable to a specific class of molecular design problems [70], [80].

### 2.3.4. Decomposition Methods

The combinatorial nature of the design space in CAMD coupled with nonlinearities in property models render many CAMD problems unamenable to optimization algorithms. A well-known attribute to the success of solving many challenging combinatorial optimization problems is the application of decomposition methods [81]. This is also true for the CAMD MINLP problem, for which many problem-specific decomposition approaches have been developed. This family of methods decomposes the original optimization problem into easier subproblems that are solved either sequentially or iteratively. In the sequential approach, the optimization problem is constructed by the sequential addition of constraints of increasing complexity. On the other hand, the iterative approach produces feasible solutions by approximating the constraints of subproblems and iterating between the subproblems to compute upper and lower bounds on the original problem [4]. These approaches make the MINLP problem easier to solve and often provide guarantees on the optimality of convex problems [82]. While most decomposition approaches in the literature have been dedicated to the more complex problems of the mixture and integrated process/product design [83]–[85], the single-molecule design remains a starting point and a building block for the other classes of problems [67], [82]. Although decomposition frameworks have achieved success in solving problems of significance, the interdependence between molecular design level and product/process design level is often oversimplified [86]. In practice, there is a complex nonlinear relationship between molecules and process performance, with the optimal solution corresponding to an intricate trade-off between different molecular properties and process variables. In such an interlinked system, sequential or iterative decomposition of molecular design problems may be suboptimal [83], [87]. A more detailed relationship between molecular design (solvent) and process design (crystallization-based chemical process) has been reported recently [88].

## 2.4. CAMD Tools and Applications

### 2.4.1. CAMD Tools

There exist several databases and software packages for CAMD [89]. A simplistic approach to providing fast solutions to single-molecule design problems is database search with listed target properties or with the help of efficient QSPR methods [90]. There is an increasing number of chemical databases that contain molecular structures in machine-readable formats with some molecular and property data. Some examples are GDB-17 [91], the largest publicly available database with more than 1.6 billion compounds, and PubChemQC [92], which has around four million compounds with property data generated from DFT calculations. As database search solutions may be suboptimal, violate other process and mixture constraints, and exclude novel molecules, a superior class of tools is referred to as model-based design in which a problem-specific mathematical optimization model is formulated and solved. Leading packages include ProCAMD [90], AMODEO [56], and OptCAMD [34]. This class of tools is highly versatile to a large variety of problems, to which it provides globally optimal solutions, but they still suffer from some drawbacks related to the size and complexity of molecules, the lack of property models for some molecules and products, and the accuracy of available property models [89].

### 2.4.2. Applications

A vast array of various CAMD applications has been growing over the past few decades for different problem scales, property estimation methods, and solution techniques. Progressively, applications have played an integral role in the development of QSPR models, solution algorithms, and frameworks. Here, we give an overview of some popular applications that have attracted much interest from the CAMD community. It is worth noting that CAMD applications have been focused on major applications in the chemical, environmental, and pharmaceutical industries, such as liquid-liquid extraction, $CO_2$ capture, polymer design, and reaction and product solvents, etc.

The development of single-molecule design applications started in the early 1980s, initially with generate-and-test approaches that were later extended to include improved screening methods and mathematical programming. In an early example of such advances, a decomposition-based generate-and-test and mathematical programming approaches were to liquid-liquid extraction [93], [94]. The efficiency of mathematical programming-based and metaheuristic frameworks was illustrated in polymer design case studies [64], [80]. Later work further explored the more challenging variants of mixture design. Examples of initial work in mixture design considered the feasibility problem for the solvent design [4], [84], [95], followed by incorporating equations of state into optimization models for refrigerant mixtures design [62]. Notable, recent efforts developed decomposition methods in conjunction with derivative-free optimization and quantum chemistry-based COMSO/COMSO-RS thermodynamics, demonstrating their efficiency with instances in solvent design, liquid-liquid extraction, and reactions solvents [40], [67], [79].

Although most CAMD methods and applications focus on single-molecule and mixture design, the solution to these classes of problems are usually integrated into a process or a final product. As such, an explicit relationship between a molecule and a process/product is necessary to optimize the process/product performance. Some of the few resources that developed applications for the integrated problem predominantly used decomposition methods [83], [86], [96]–[98]. Also, integrated product/process design applications using mathematical programming [67], [99], [100], and metaheuristics [101]–[103] are present. A brief list of references classified by design level and major application is given by Table 1.

**Table 1.** A brief list of popular CAMD applications classified by design level and application.

| Design Level | Application | References |
|---|---|---|
| **Single-Molecule** | Liquid-liquid extraction | Austin et al. [40], Brignole et al. [104], Diwekar and Xu [71], Gani and Brignole [93], Gani et al. [105], Gebreslassie and Diwekar [106], Harper et al.[107], Harper and Gani [55], Karunanithi et al. [82], Kim and Diwekar [108], Marcoulaki and Kokossis [109], Odele and Macchietto[94], Ourique and Telles [110], Scheffczyk et al. [111] |
| | Polymer design | Brown et al. [112], Camarda and Maranas [60], Eslick et al. [77], Maranas [64], Pavurala and Achenie [113], Venkatasubramanian et al. [74], [80], Zhang et al. [61]. |
| | Reaction solvents | Wang and Achenie [114], Gani et al. [115], Folic et al. [116], [117], Struebing et al. [118], Zhou at al. [119] |
| | Refrigerant design | Churi and Achenie [120], Duvedi and Achenie [57], Gani et al. [105], Joback et al. [121], Marcoulaki and Kokossis [109], Ourique and Telles [110], Sahinidis et al. [59], Samudra and Sahinidis [122] |
| **Mixture** | Solvent design | Austin et al. [40], [67], [79], Buxton et al. [99], Conte et al. [123], Duvedi and Achenie [62], Gani and Fredenslund [95], Karunanithi et al. [82], Klein et al. [84] |
| **Integrated Process/ Product** | $CO_2$ Capture | Bardow et al. [83], Burger et al. [100], Gopinath et al. [86], Lampe et al. [96], Pereira et al. [124], Stavrou et al. [97] |
| | Gas absorption | Buxton et al. [99], Papadopoulos and Linke [103], Bommareddy et al. [98], Zhou et al. [72] |

## 3. Deep Learning for Molecular Design

An ever-increasing volume of research studies has been reported in applying deep learning models to molecular generation and property prediction of molecules. To address the CAMD problems, generative models are used in conjunction with predictive QSPR models that relate learned feature representations of molecular descriptors to target chemical, physical, or biological properties of structures. In addition to beating other machine learning methods in property predictions, deep learning has recently demonstrated the capability to produce property predictions comparable to DFT calculations [18], [20]. Thus, the focus of this section is more along the lines of generative models and their synergy with predictive models for molecular design and optimization. A high-level description of popular architectures for generative models in the literature is displayed in Figure 2.

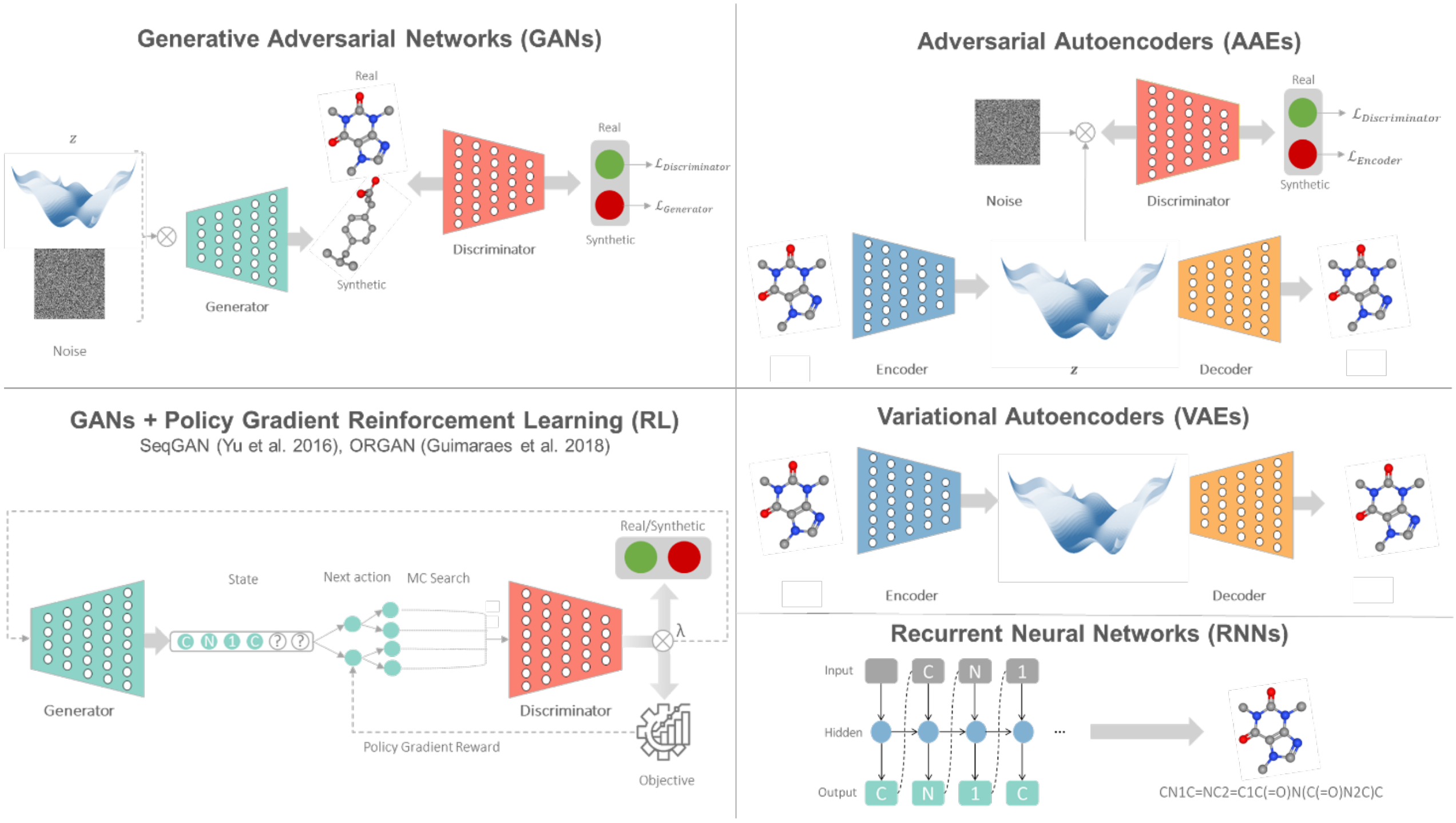

**Figure 2.** High-level diagrammatic description of common architectures in the literature.

The section starts by reviewing molecular representations used as input to models. The core of the section is devoted to developing a comparative and contrastive summary of published generative models and methods that hold the most potential for integrating molecular generation with property prediction. A brief discussion on evaluation metrics and benchmarking platforms concludes the section.

## 3.1. Molecular Representations

Central to the learning process is the digital encoding of an expressive molecular structure representation that captures the structural information about a molecule using a specific format and definite rules. The selection of the method that translates the structural information into a machine-readable format is termed featurization or feature engineering. In the context of molecular design and discovery, popular representations fall into two classes: 2-dimensional and 3-dimesnsional. The suitable selection of a molecular representation requires insights into the specific problem and the intended machine learning algorithm. However, the selection of the best-performing representation for a learning task or algorithm is not always apparent, remaining an open research question in cheminformatics [125]. Figure 3 illustrates a few commonly used molecular representations including: coulomb matrix [126], fingerprints [127], InCHI [128], SMILES [129], Junction-Tree [130], and molecular graph convolution [131].

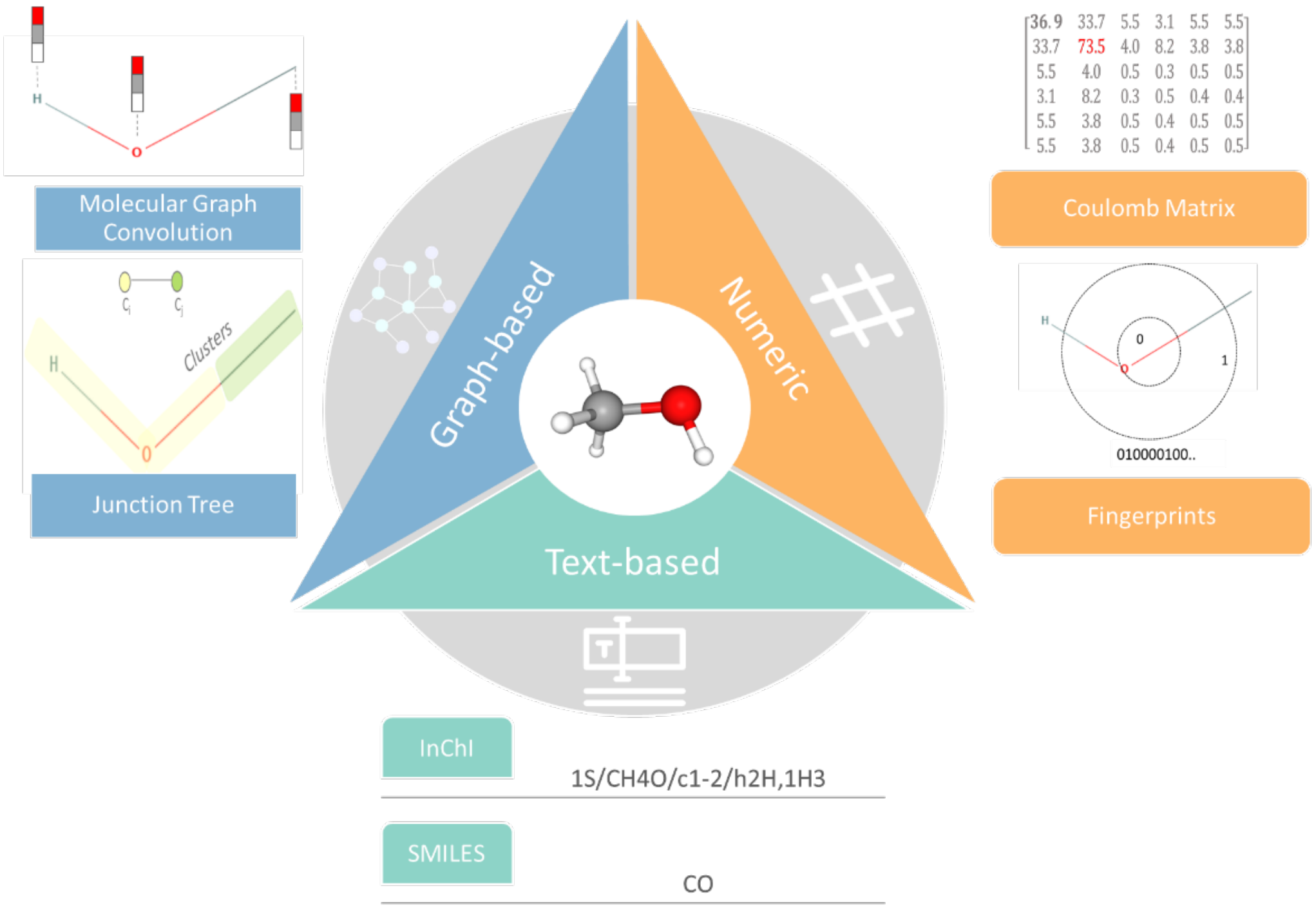

**Figure 3.** A few commonly applied molecular representations in deep learning for methanol.

Before shedding light on the different molecular representation methods, it is imperative to highlight the key invariances that certain representations capture: permutation, translational, and rotational. Permutation invariance allows the representation to remain unchanged under reordering or permutation of atoms. Representations unaltered by translational and rotational operations in the chemical space satisfy the other two invariances [132]. These invariances can be reduced by two desirable properties of machine-readable representations: uniqueness and invertibility. Whereas the uniqueness property is fulfilled when any molecule can be represented by a single expression, invertibility goes in the opposite direction and is satisfied with a one-to-one mapping between representations and their corresponding molecule. Since the generative power of deep learning is fundamental to the problem of molecular design, most implementations use invertible representations [133].

There is sheer size of publically available databases of compounds for use by researchers undertaking molecular design. Table 2 provides a list of public databases of molecules, and their size and downloadable representations. It is noteworthy that the molecular representations provided by these public databases can be validated and translated into different molecular representations using several software tools and packages such as RDkit [134], OpenBabel [135], and OpenEye [136]. The sources of the property values included in these databases range from text-mined experimental data and pseudo-experimental DFT calculations, to predicted values with varying degrees of confidence. Thus, it is critical to maintain a degree of skepticism about the accuracy of property values, and factor in uncertainties and underlying sources. Popular molecular 2D and 3D representations and their relative merits are reviewed in the following subsections.

**Table 2.** A list of representative publicly available chemical databases including descriptions of their content, the number of compounds, and available molecular representations.

| Database | Content | Size | Representations | Ref. |
|---|---|---|---|---|
| **GDB-17** | Enumeration of small organic molecules up to 17 atoms of C, N, O, S, and halogens. | 166,400,000,000 | SMILES | [91] |
| **ZINC15** | Commercially-available compounds. | >750,000,000 | SMILES, InChI, InChI Key, 2D SDF, 3D SDF | [137] |
| **SureChEMBL** | Compounds extracted from patent documents. | 17,000,000 | SMILES, InChI Key | [138] |
| **eMolecules** | Commercially-available compounds. | 5,900,000 | SMILES, 2D SDF | [139] |
| **PubChemQC** | Compounds with quantum chemistry estimated property based on DFT methods. | >3,500,000 | SMILES, InChI | [92] |
| **ChEMBL** | Bioactive molecules with drug-like properties. | 1,900,000 | SMILES, InChI, InChI Key, 2D SDF | [140] |
| **SuperNatural** | Natural molecules with physicochemical properties and toxicity class. | 325,508 | SMILES | [141] |

#### 3.1.1. 2D Representations

2D representation-based learning-based molecular design models have enjoyed remarkable success, although the use of such representations entails the loss of conformational and bond distance information [142]. As described in the previous section, the use of the chemical graph theory to represent molecules is a natural way to apply the principles of bonding and chemistry valence rules, yet the most popular representation in data-driven molecular design literature is the string-based SMILES [128], [133]. Similar to most graph methods, SMILES do not correspond to unique representations. This non-uniqueness can be resolved by converting standard SMILES into its canonical form. However, such a feature has been proven to be quite useful as a data augmentation technique for machine learning [143]. Another common string-based representation is the IUPAC International Chemical Identifier (InChI) [144]. It is reported that the use of InChI in a variational autoencoder-based approach led to substantially better performance than SMILES in the learning task due to the more complex syntax that involves counting and arithmetic [25]. Recent work showed that a more meaningful chemical latent space is learned by translating between InChI and SMILES using variational autoencoders [145]. A typically preprocessing step of string-based representations includes converting them to one-hot encoding or molecular descriptors [25], [146]. In molecular generation, achieving a perfect rate of valid reconstruction for SMILES from the latent space remains an open problem, despite the availability of well-developed deep learning architectures for various sequence generation applications, such as music synthesis and natural language processing [142].

Many other graph representations, such as 2D images, tensors, and MACCS keys, have been applied to molecular generation with varying success rates. Inspired by the success in image classification by Google's Inception-ResNet deep convolutional neural network (CNN) [147], a deep CNN, Chemception, was developed using image data of 2D drawings of molecules encoded in single-channel greyscale for property prediction [148]. Also, 2D image embedding combined with standard SMILES were used as an input to a heteroencoder, resulting in a better latent space compared to SMILES and canonical SMILES [143]. Operating in the space of graphs through the use of vectors and tensors to encode adjacency and connectivity information has been shown to be a promising alternative. Several successful implementations of tensor-based representations have achieved perfect reconstruction validity and high rates of novelty for small molecules [149], [150]. It should be noted that more compact representations are applicable for small datasets and specific classes of molecules in molecular generation. Gene expressions were generated using the 166bit MACCS with a valid decoding rate below 10% [151], [152]. It is also worth noting that such compact representations are better suited for regression tasks with many successful implementations in property prediction using bag of bonds, fingerprints, Coulomb matrices, and molecular descriptors [146], [153]–[155].

### 3.1.2. 3D Representations

Owing to the variance of 3D representations with translational, rotational, and permutation operations, describing molecules in the 3D space might not be best suited for generative models [142]. An approach to express molecules in the 3D grid space is achieved using voxels with localized channels for distinct types of atoms and nuclear charge information. This approach encounters the challenges of high dimensionality and sparsity, leading to high complexity and poor performance of the generative model [133]. To ease the computational effort and avoid sparsity, a wave transformation-based approach was proposed, replacing each atom with concentric waves diverging from its center using the wave transform kernel [156]. Another key contribution on this front is a novel, but not invertible, tensor field neural networks framework that is equivariant to 3D rotations, translations, and permutations of 3D coordinates of points [132], [157]. The challenges and limitations of this class of representation underline the need for research work that develops efficient invertible 3D representations for molecular design.

## 3.2. Deep Learning Architectures for Molecular Design

### 3.2.1. Recurrent Neural Networks (RNNs)

RNN is a prevalent class of deep learning models for sequence generation, constituting a building block of other generative deep learning architectures. Equipped with parameter-sharing and graph-unrolling schemes, this architecture is capable of mapping relationships and dependencies between molecular character sequences of arbitrary lengths by introducing a notion of time or order to the model [158]. The arbitrary-length sequence is dealt with by incorporating a hidden state whose activation is dependent on that of the previous element in the molecular sequence. For generative models in molecular design, an RNN produces a probability distribution of the next character of the sequence based on its current hidden state, where the last character in the sequence is a special value. For a given sequence, $s = (s_1, s_2, \ldots, s_t)$, the probability of the sequence can be expressed as:

$$p(s_1, s_2, \ldots, s_t) = p(s_1)\, p(s_2|s_1)\, p(s_1|s_1, s_2) \ldots p(s_t \mid s_1, s_2, \ldots, s_{t-1}) \tag{10}$$

For $n$ training sequence, a model parametrized by $\theta$ learns the probability distribution over the sequences by maximizing the log-likelihood on the sequence space:

$$\max_{\theta} \frac{1}{N} \sum_{n=1}^{N} \sum_{s=1}^{S} \log p(s_t^n \mid s_1^n, s_2^n, \ldots, s_{t-1}^n; \theta) \tag{11}$$

The vanishing and exploding gradient problems tend to arise when training RNNs to capture long-term dependencies, making the application of gradient-based optimization unsuitable [159]. A dominant approach to alleviate this issue is the development of more sophisticated activation functions that involve gating mechanism for element-wise nonlinearity following affine transformations [160]. Attempts in these directions have resulted in two types of recurrent units that have been shown to perform well for sequence modeling: the long short-term memory (LSTM) unit and the more recent gated recurrent unit (GRU) [161], [162]. For the task of molecular generation, several papers have demonstrated the potential of LSTM-based RNNs, reporting generated structures validity rates as high as 97.7% [143], [163], [164]. A remarkable outcome was also achieved using a three-layer GRU-based RNN with 94% of the SMILES generated by the prior corresponding to valid molecules [165]. Despite such successes, it is suggested that the introduction of an external stack is essential for the generation of valid SMILES owing to the limited counting ability of LSTM and GRU units [26]. Instead, their work employs a stack-augmented RNN (Stack-RNN) that defines a new memory cell on top of a GRU recurrent units to better infer algorithmic patterns [166]. Under the Stack-RNN generative model, 95% of SMILES were found to be valid compared to an 86% validity rate with the same RNN architecture excluding the stack memory. Such differences in validity rates may be resultant from a variety of factors, including different databases, sampling approaches, and molecular validation tools.

Several property prediction models have been implemented in conjunction with SMILES-based RNN architectures. Two frameworks have compared several predictive algorithms to perform the property prediction task, reporting Support Vector Machine (SVM) and Gradient Boosting Trees (GBT) as the best classifiers of Dopamine receptor $D_2$ and half-maximal inhibitory concentration ($pIC_{50}$) properties, respectively [164], [165]. Deep learning models are also commonly applied for the predictive task. With SMILES string as an input vector, the ReLeaSE framework uses a multilayer neural network connected to an embedding recurrent LSTM layer for predicting four properties [26]. Generative and predictive models are coupled under a reinforcement learning system for optimizing target properties of generated molecules, which will be covered in the Reinforcement Learning subsection.

### 3.2.2. Autoencoders

Autoencoders (AE) is a multilayer neural network that is trained to recover its input as its output by means of an internal hidden layer describing the code of the input. Connected through the code layer, AEs consist of two neural networks, serving as an encoder and a decoder. In molecular generation, the encoder translates each representation, SMILE, into a fixed-dimensional vector, while the decoder does the opposite stochastic mapping operation [158]. These two networks aim to learn the identity function, whereas the role of the latent representation is to induce the networks to learn a reduced representation that allows for capturing the most salient and descriptive information of the representation [167].

By exploring theoretical relationships between the latent space and AEs [168], the formulation of variational autoencoder (VAE) was proposed in 2014 and first applied to molecular design in 2017 [167], [169]. Later developments in the VAE formulation included new models for semi-supervised VAE (SSVAE), which incorporates approximate Bayesian inference with advances in variational methods to improve the quality of the generative modeling approach [170]. Another dominant class of probabilistic autoencoders is the adversarial autoencoder (AAE), which employs the GAN framework as a variational inference algorithm for latent variables [171], [172]

#### 3.2.2.1. *Variational Autoencoders*

VAEs are latent variable models comprised of latent variables $z^{(i)}$ that are drawn from a prior $p(z)$ and fed into a decoder $p_\theta(x|z)$. The central idea behind VAEs is to sample values of the latent variable that are likely to have generated *x* and only use such values in the computation of the probability of *x* in the training set. To perform this task, a new encoding function $q(z|x)$ is needed to provide a distribution over $z$ values that are likely to return $x$. The training procedure is done by maximizing the variational lower bound given by Eq. (12), which allows for the loss function to be written as in Eq. (13).

$$\log p_\theta(x) \geq \mathcal{L}(\theta,\phi;x) = \mathbb{E}_{z\sim q\phi(z|x)}\left[-\log q_\phi(z|x) - \log p_\theta(\mathrm{x},\mathrm{z})\right] \tag{12}$$

$$\mathcal{L}(\theta,\phi;x) = \mathbb{E}_{z\sim q\phi(z|x)}\left[\log \mathrm{p}_\theta(x|\mathrm{z})\right] - D_{KL}\left(q_\phi(z|x) \,||\, p_\theta(z)\right) \tag{13}$$

where the first term in the right-hand side is the variational lower bound and the Kullback–Leibler divergence term serves as a regularizer.

VAEs have proven to possess a very powerful generalization ability due to the stochasticity encoded within the learning method, describing molecules as continuous probability distributions instead of discrete fixed points. This is particularly advantageous for the problem of molecular design, as the probabilistic nature of the formulation forces the latent space to have robust and diverse representations. Further, this feature is especially useful for the construction of open-ended spaces for chemical compounds which allow for not only the generation of new molecules, but also interpolating between existing ones [25].

Early VAE implementations to molecular generation have suffered from low valid rates of SMILES outputted by the decoder, suggesting underlying issues within the latent space [25], [169]. Grammar VAE (GAVE) and character VAE (CVAE) have reported valid decoding rates of 0.7% and 7.2%, respectively [169]. Another VAE model resulted in a decoding rate of 4% for randomly sampled points from the latent space and a rate that ranges from 73% to 79% when sampling around known molecules. The introduction of syntax-directed translation to the SMILES used as an input to the VAE increased the valid decoding rate to 43.5%. VAEs with graph representations as input have shown much better performance with several sources reporting up to 100% validity of generated molecules using hypergraph grammar, constrained graphs, and graph-to-graph translation [173]–[175].

#### 3.2.2.2. *Supervised/Semi-Supervised Autoencoders*

A unique advantage that SAE/SSAE offer is the ability to connect the property prediction into the molecular generation paradigm and conditional sampling [176]. This type of architecture offers a unique advantage to the CAMD problem by working with partially labeled datasets in which target properties are available for a subset of molecules. To incorporate chemical properties as an output variable $y$, a generative semi-supervised model can be adopted under the assumption that the target properties have a Gaussian distribution [170], [176]. Since the exact form of the posterior distribution is intractable, an approximate form is used to estimate the posterior distributions over $y$ and $z$. The variational lower bound for labeled data are given as follows:

$$\log p_\theta(x,y) \geq -\mathcal{L}(x,y) = \mathbb{E}_{q\phi(z|x,y)}\left[\log p_\theta(\mathrm{x}|\mathrm{y},\mathrm{z}) - \log q_\phi(z|x,y) + \log p_\theta(\mathrm{y}) + \log p(z)\right] \tag{14}$$

For molecules with an unobserved label, a target property, *y*, is treated as a latent variable over which posterior inference is performed, resulting in the variational bound below:

$$\log p_\theta(x) \geq -\mathcal{U}(x) = \mathbb{E}_{q\phi(y,z|x)}\left[\log p_\theta(\mathrm{x|y,z}) - \log q_\phi(y,z|x) + \log p_\theta(\mathrm{y}) + \log p(z)\right] \tag{15}$$

Under this class of models, it is possible to generate molecules with target properties through sampling from a conditional generative distribution given a set of desired properties, obviating the need for any optimization procedure. A leading result was reported by an SSVAE model that achieves a >99% validity rate of generated SMILES with more than 92% of SMILES being unique [176]. It is noteworthy that the same concepts with slight variations can be followed for constructing a semi-supervised AAE (SSAAE), which was adopted recently for generating molecules that optimize the half-maximal inhibitory concentration [177]. We anticipate that this class of models will be most promising in constructing an extensive chemical latent space owing to its ability for learning meaningful disentanglements of data [178].

#### 3.2.2.3. *Adversarial Autoencoders*

AAEs are similar to VAEs with the only difference being associated with the regularization term. AAE replaces the Kullback–Leibler divergence term with an adversarial training procedure to impose a prior distribution on the code vector of the AE [171]. Building on the GAN framework, the adversarial process simultaneously trains two neural networks: a generative model, *G*, that captures the data distribution and generates new samples, and a discriminative model, *D*, that distinguishes between the prior distribution and the encoding produced by *G* [172]. The overall loss function combining the adversarial loss for the discriminator and the reconstruction loss can be expressed as:

$$\mathcal{L}(\theta,\phi;x) = -\mathbb{E}_{x\sim p_d}\left[\log D\left(q_\Theta(z|x)\right)\right] - \mathbb{E}_{x\sim p_z}\left[\log\left(1 - D(z)\right)\right] + \Sigma - \mathbb{E}_{x\sim p_d}\left[\log p_\theta\left(x \mid q_\phi(z|x)\right)\right] \tag{16}$$

Only a few implementations of AAE have been reported in the literature with severe difficulties in the adversarial training process, even for small datasets [179]. Common across all AE-based models, differentiable predictive models can be trained for property prediction using continuous latent representations that correspond to a subset of molecules. This enables gradient-based optimization methods to navigate the chemical design space and move towards the direction of optimizing a given objective of target properties, avoiding the complications induced by the discrete nature of the chemical space [25]. A special algorithm for optimized property-oriented decoder was able to identify molecules with property values 121% higher than Bayesian optimization and reinforcement learning [180]. Similar results are reported for the molecular hypergraph variational approach (MHG-VAE) for molecular optimization when the target function evaluations are limited, outperforming reinforcement learning and GAN-based Graph Convolutional Policy Network (GCPN) in terms of computational cost [173], [181].

### 3.2.3. Generative Adversarial Networks (GANs)

GAN constitute another class of generative models that constructs a latent space to simplify molecular representations into a compressed representation shared across the molecular domain. Serving as the theoretical basis of the idea of adversarial training introduced earlier in AAE, in molecular design, the framework pits a generative neural network against an adversarial neural network that aims to discriminate between the generated molecules distribution and the original molecules used for training [172]. For a generative model, *G*, and a discriminative model, *D*, the theoretical minimax objective function is expressed as:

$$\min_{G}\max_{D} V(D,G) = \mathbb{E}_{x\sim p_d(x)}\left[\log D(x)\right] + \mathbb{E}_{z\sim p_z(z)}\left[\log\left(1-D\left(G(z)\right)\right)\right] \quad (17)$$

Optimizing the above function is done iteratively and with some alterations to the function to provide better gradients early in learning [172]. Also, GANs offer much more flexibility than AE in terms of defining the objective function using the Jensen-Shannon divergence, f-divergences, or a combination of them [182]. However, training GANs is characterized by instability and delicacy of parameters, as it requires locating a Nash equilibrium of a non-convex game with continuous, high-dimensional parameters [182], [183]. Though GANs remains a current research focus, improvements to training stability and sample quality have been proposed and implemented for molecular generation, such as Wasserstein GAN (WGAN) and SeqGAN [183], [184].

Several extensions to the GANs framework have been adopted to generate molecules with limitations associated with the generation of valid SMILES representations. An Objective-Reinforce GAN (ORGAN) model has a molecule validity rate that fluctuates from 0.01% to 99.8%, attributing such outcome to the sparsity and roughness of the chemical space that leads to a poor generator performance [185]. Among other issues in adversarial training settings, the vanishing gradient problem arises from the convergence of the minimax objective to a saddle point, where the gradient of the generator vanishes given a discriminator that perfectly labels the real and generated data. In an attempt to escape saddle points, an RNN-based differentiable neural computer (DNC) with access to an external memory cell was implemented, leading to a 76% rate of valid SMILES [186], [187]. A parallel validity rate was achieved by an adversarial threshold neural computer (ATNC) structure, which acts as a filtering agent between the generator network and the discriminator and reinforcement learning networks [188]. A more effective and stable approach combined the SeqGAN framework for sequential data with WGAN as a loss function, obtaining a valid SMILES rate of 80.3% [189].

Graph-based GAN implementations demonstrated their ability to offer near-perfect validity of generated molecules. An implicit GAN and reinforcement learning framework for molecules had success in achieving high rates of validity running up to 99.8%. However, this approach attains a maximum rate of unique molecules of 3.2%, alluding to the issue of mode collapse, wherein the generator rotates through a small set of molecules that seem most plausible to the discriminator [150], [190], [191]. Further, on the property optimization front, a superior performance was reported under a graph generation policy network (GCPN) [181]. Yet, two common issues are often encountered under graph-based GANs: the limited diversity in generated molecules and handling the graph isomorphism problem.

### 3.2.4. Reinforcement Learning

Reinforcement learning has been applied in the literature to generate molecules with desirable properties and fine-tune the performance of the GAN generator [189]. The framework requires a learning agent to learn a stochastic policy that maps states to action probability vectors, so as to maximize a reward [192]. Commonly applied reinforcement learning-based approaches in CAMD include the two main model-free approaches to learning an optimal policy: methods based on policy search and value functions methods [165]. Reinforcement learning embodies a natural environment for a more powerful generate-and-test approach to the CAMD problem.

SMILES with optimized properties were generated by a partially observable Markov Decision Process (MDP) RNN using policy search methods. In this approach, the use of policy search methods is argued to be a more intuitive approach as it possesses the ability to start with a prior sequence model as an initial policy, requiring shorter episodes and leading to an optimal policy [165]. A comparative study for optimizing molecules towards property targets was carried out between several policy-based methods: Proximal Policy Optimization (PPO) and REINFORCE, a hybrid advantage actor-critic method (A2C), and Hillclimb-MLE [163], [193]–[195]. The study ranked the reinforcement learning methods based on 19 benchmarks for molecular generation and design, suggesting that the Hillclimb-MLE method outperformed the rest given sufficient computational times and sample evaluations [163]. In contrast to model-free methods, a few model-based reinforcement learning methods coupled with adversarial training have been implemented. A notable model-based implementation is ORGAN, discussed in the GAN section [185]. The demonstrated success of the ORGANIC framework paved the way for further efforts to build on the framework and its use for benchmarking new GAN-based reinforcement learning approaches [150], [187], [188], [190].

**Table 3.** A detailed list of published molecular generation and optimization works. **Acronyms**: *(AAE) Adversarial Autoencoder; (ATNC) Adversarial Threshold Neural Computer; (BI) Bayesian Inversion ; (BL) Bayesian Learning; (BNN) Bayesian Neural Network; (BO) Bayesian Optimization; (CGVAE) Conditional Graph Variational Autoencoder; (CVAE) Conditional Variational Autoencoder; (DNN) Deep Neural Network; (DRD2) Dopamine Receptor D2 ; (GAN) Generative Adversarial Network; (GBT) Gradient Boosted Trees; (GCPN) Graph Convolutional Policy Network; (GGNN) Gated Graph Neural Networks; (GP) Gaussian Process; (GVAE) Graph Variational Autoencoder; (HBA) number of hydrogen acceptor; (HBD) hydrogen bond donor; (HL) HOMO-LUMO gap ; (JAK2) Janus kinase 2; (JT) Junction Tree; (LogP) Partition coefficient; (MHG) Molecular Hypergraph Grammar ; (MW) Molecular weight; (pIC50) half maximal inhibitory concentration; (POD) Property Oriented Decoder; (QED) Quantitative Estimate of Drug- likeness; (RL) Reinforcement Learning ; (RNN) Recurrent neural network; (RO5) Lipinski's rule of five; (SA) synthetic accessibility; (SGP) Sparse Gaussian Process; (SSAAE) Semi-supervised Adversarial Autoencoder; (SSVAE) Semi-supervised Variational Autoencoder; (SVM) Support Vector Machine; ($T_m$) Melting Temperature; (TPSA) topological polar surface area; (U) Internal Energy; (VAE) Variational Autoencoder.*

| Generative Model | Representation | Predictive Model | Predicted/Optimized Features | Ntrain | Database | Ref. |
|---|---|---|---|---|---|---|
| **RNN** | SMILES | RNN/RL | MW, LogP, HBD, HBA | 1,735,442 | ChEMBL | Neil et al. [163] |
| **RNN** | SMILES | DNN/RNN | Tm, logP, pIC50, JAK2 | 1,500,000 | ChEMBL21 | Popova et al. [26] |
| **RNN** | SMILES | SVM | DRD2 | 1,500,000 | ChEMBL | Olivecrona et al. [165] |
| **RNN** | SMILES | GBT | pIC50 | 1,400,000 | ChEMBL | Segler et al. [164] |
| **BL** | Fragments | BI | HL, U | 60,000 | PubChem | Ikebata et al. [196] |
| **VAE** | SMILES | DNN | LogP, HBD, HBA, TPSA | 500,000 | ZINC | Lim et al. [197] |
| **VAE** | SMILES | DNN/GP | LogP, QED, SA | 358,000 | QM9/ZINC | Gómez-Bombarelli et al. [25] |
| **VAE** | Graph (embedded vectors) | BO/ RL/ POD | LogP, QED | 20,000 | ZINC/QM9 | Samanta et al. [180] |
| **VAE + BNN** | SMILES | BNN/BO | LogP, QED, SA, #rings | 249, 456 | ZINC | Griffiths and Hernández-Lobato [198] |
| **AAE (druGAN)** | MACCS | NA | Predefined anticancer properties | 6,252 | HMS LINCS | Kadurin et al. [179] |
| **CGVAE** | Graph | GGNN | QED | 250,000 | QM9/ZINC/XEPDB | Liu et al. [174] |
| **CVAE/ GVAE** | SMILES | BO | LogP | 250,000 | ZINC | Kusner et al. [169] |
| **JT-VAE** | Graph | SGP | LoP, SA, #cycles | 250,000 | ZINC | Jin et al. [130] |
| **JT-VAE+GAN** | Graph | GNN | LogP, QED, DRD2 | 250,000 | ZINC | Jin et al. [175] |
| **MHG-VAE** | Graph | GP | LogP, SA, #cycles | 250.000 | ZINC | Kajino [173] |
| **SSVAE** | SMILES | RNN | MW, LogP, QED | 310,000 | ZINC | Kang and Cho [176] |
| **SSAAE** | SMILES | Disentanglement | logP, SA | 1,800,000 | ZINC | Polykovskiy et al. [177] |
| **GAN** | SMILES | RL | LogP, SA, QED | 5,000 | ZINC | Guimaraes et al. [189] |
| **GAN (ORGAN)** | SMILES | RL | MW, LogP, TPSA | 15,000 | ZINC/ChemDiv | Putin et al. [187] |
| **GAN** | SMILES | RL | Tm, QED, RO5 | 15,000 | GDB-17/ZINC | Sanchez-Lengeling et al. [185] |
| **GAN** | Grammar SMILES | NA | Predefined transcriptomic profile | 19,800 | L1000 CMap | Méndez-Lucio et al. [152] |
| **GAN (ATNC)** | SMILES | RL | MW, LogP, TPSA | 15,000 | ChemDiv | Putin et al. [188] |
| **GAN (Cycle)** | Graph | RL | LogP, rings | 250,000 | ZINC | Maziarka et al. [190] |
| **GAN** | Graph | RL | LogP, SA, QED | 133,885 | QM9 | Cao and Kipf [150] |
| **GAN (GCPN)** | Graph | RL | LogP, QED, MW | 250,000 | ZINC | You et al. [181] |

## 3.3. Benchmarking in Molecular Design

Despite the extraordinary advances in deep learning models, their performance is not directly realized from CAMD results, especially for generative models where no clear-cut exists for comparing generated molecules. Many researchers have observed that the emphasis on yielding impressive empirical results might not have been matched with a parallel emphasis on empirical rigor [199]–[201]. It is argued that this creates a bias towards implementing newer approaches that are claimed to outperform classical approaches [199]. Two large-scale studies on evaluating the performance of new generative approaches found no evidence that newer approaches consistently score better than original formulations with sufficient hyperparameter tuning and random restarts [133], [201], [202]. As such, empirical rigor and standardized benchmarks and datasets are critical to triggering progress towards better CAMD models and algorithms.

Testing the promising new generative models on consistent tasks and comparing their performance to classical methods is vital to accelerate the pace of progress toward automated molecular discovery and optimization. A molecular design framework should be assessed on two elements: the characteristics of generated molecules and the optimization of an objective function of target properties. In the literature, generative models have been widely assessed based on certain properties, such as LogP, QED, or SA, making the assessment process difficult and uninformative [203]. While many metrics and reward functions are presented and applied in the literature, two deep learning benchmarking platforms for molecular design have been developed for the evaluation of models and algorithms in a controlled setting: MOSES and GuacaMol [204], [205].

### 3.3.1. MOSES

MOlecular SEtS (MOSES) offers a combined implementation of a benchmarking platform that consists of data preprocessing tools, a standard database, and evaluation metrics, along with state-of-the-art molecular generation models. In contrast to GaucaMol, MOSES is solely concerned with evaluating the task of molecular generation. To compare a generated set of molecules, *G*, against a reference set of molecules, *R*, taken from the training set, the platform defines five main evaluation metrics. Most of the metrics are similarity measures - fragment similarity (Frag), scaffold similarity (Scaff), Fréchet ChemNet Distance (FCD), nearest neighbor similarity (SNN), and the internal diversity ($IntDiv_p$). The platform also includes auxiliary metrics that are commonly used for small molecule drug discovery, including MW, LogP, SA, QED, and natural product-likeness score (NP).

Frag computes the cosine distance between the fragment frequencies of the generated and reference molecules, $f_G$ and $f_R$, using the BRICS algorithm [206] to decompose molecules into chemical fragments. Similarly, Scaff calculates the cosine distance between the scaffolds frequencies, *s*, which are produced by implementing the Bemis-Murcko algorithm to remove the side chain atoms [207].

The SNN metric is expressed as the average of the Tanimoto distance between a fragment representation of a generated molecule, $m_G$, and its nearest neighbor from the reference set, $m_R$:

$$SNN\left(G,R\right)=\frac{1}{\left|G\right|^{2}}\sum_{m_G\in G}\max_{m_R\in R}T\left(m_G,m_R\right) \tag{18}$$

Another metric that uses the Tanimoto distance to assess the diversity of generated molecules is the internal diversity metric proposed by Benhenda [208]. The internal diversity is different from other similarity metrics in that it offers insight into the diversity of the generated molecules, allowing to detect flows in the generative model as mode collapse.

$$IntDiv_p\left(G\right)=1-\sqrt[p]{\frac{1}{\left|G\right|^{2}}\sum_{m_1,m_2\in G}T\left(m_1,m_2\right)^{p}} \tag{19}$$

Similar in principal to Fréchet Inception Distance [209], FCD computes the distance between the distribution of molecules in the dataset and generated molecules using: the activations of the penultimate layer of the “ChemNet” LSTM-RNN, mean vectors, $\mu$, and covariance matrices, $\Sigma$, for each distribution [203].

$$FCD\left(G,R\right)=\mu_G-\mu_R{}^{2}+Tr\left(\Sigma_G+\Sigma_R-2\sqrt{\Sigma_G\Sigma_R}\right) \tag{20}$$

### 3.3.2. GuacaMol

GuacaMol outlines two categories of quantitative benchmarks for molecular design models: distribution-learning and goal-directed. The distribution-learning category uses five benchmarks to quantify the quality of a generative model trained to reproduce a distribution of molecules from the training set. On the other hand, goal-oriented benchmarks employ robust and simple scoring functions to disentangle the selection of a good scoring function from the problem of molecular optimization. Under goal-oriented benchmarks, the objective function represents a combination of different molecular features, such as physiochemical properties and structural features [205]. The two classes of benchmarks are evaluated separately to better analyze the performance of a molecular design framework. However, such evaluation rests on the assumption that there is no one-to-one correspondence between the two tasks, and this assumption may not hold for reinforcement learning-based approaches. Even so, the benchmarking platform offers a unique quantitative route to advance the comparability of models in terms of molecular optimization. The following paragraphs provide a summary of the key metrics within the two categories of the platform.

#### 3.3.2.1. *Distribution-learning Benchmarks.*

This class of benchmarks assesses the molecule generation task using the following five benchmarks: validity, uniqueness, novelty, Fréchet ChemNet Distance, and Kullback–Leibler divergence. The validity benchmark determines the ability of a model to generate theoretically valid molecules, which can be validated using software packages, such as RDKit [134]. Further, a generative model is also evaluated based on the two overlapping benchmarks of uniqueness and novelty. The uniqueness measures the ability of the model to generate diverse molecules with no repetitions, whereas the novelty computes the fraction of molecules that are not present in the training set. Last, GuacaMol includes a common metric for distribution reconstruction, the Kullback–Leibler divergence, which captures diversity by providing a measure on how the generated molecules distribution is different from that of the training set [210].

#### 3.3.2.2. *Goal-directed Benchmarks.*

Under this category, the objective function is defined as a combination of two or more molecular features, including but not limited to: the presence of substructures, similarity to other molecules, and structural and physicochemical properties. GuacaMol establishes 20 benchmarks that compile the combinations of four main molecular features: similarity, rediscovery, isomers, and median molecules. The similarity metric quantifies the crucial form of inverse-screening in many CAMD problems, which aims to generate a molecule based on the similarity/dissimilarity to a given molecule. A closely related benchmark to similarity is the rediscovery metric, seeking to rediscover a target molecule with special importance in de novo drug design. As isomers can have very dissimilar properties, an isometry metric is included to evaluate the flexibility of a generative model in enumerating the isomers of a given molecular representation. Moreover, the median molecules metric is integrated to maximize a molecule similarity to neighboring molecules and reward encoding more molecular structures [205]. It is worth noting that this class of benchmarks is not concerned with the selection of the best scoring function, but rather it considers the complex combinations and trade-offs of molecular features.

## 4. Perspectives and Future Directions

In the preceding sections, we provided a critical survey of knowledge-based and data-driven methods and tools for molecular design and optimization. The literature presents an increasingly complex and rich array of molecular representations, QSPRs, solution methods, model architectures, algorithms, and tools. Even with many successful implementations and the sheer size of research in this direction, challenges presented in implementations point to the exigency for the molecular design community to devise multifaceted strategies and frameworks directed towards closing the loop. While the emergent deep learning methods have yet to surpass knowledge-based methods on molecular design tasks, we predict that major advancement in this application domain will come about through complex systems that integrate complex chemical knowledge with representation learning [205], [211]. Here, we offer our outlook on current approaches along with the major opportunities, challenges, and trends for this nascent class of solution methods.

### 4.1. Hybrid Knowledge-based and Data-driven Approaches

The principles of the knowledge-based and data-driven CAMD approaches are essentially different. Knowledge-based methods directly explore the chemical space, while encoding the rules for structural validity and bounding property targets [3]. On the other hand, deep learning-based CAMD methods approximate the structure of a subset of the chemical space observable as input data, constructing a latent space that preserves the required features for reconstructing the chemical space [25]. In this way, deep learning approaches are promising candidate routes for replacing and complementing the knowledge-based counterpart by side-stepping limitations associated with the complexity of molecular systems. For example, a potential alternative to complex property prediction models is the use of input-convex neural networks, which would allow for the optimization over molecules [212]. As model-based methods currently stand, frameworks and tools have long been established and used to solve problems of significance in academia and the industry [89]. Yet, several challenges lie ahead for applying knowledge-based methods to many different classes of molecules including the paucity of property data, the reliability and predictive power of property models, and the accessibility of solution strategies for multiscale design [89], [213]. Conversely, while holding the promise to transform molecular design, deep generative models are still in infancy and guided by the empirical nature of model development in machine learning. It is hence imperative to adopt and consider these challenges in the development of future data-driven or hybrid approaches.

At present, knowledge-based decomposition methods have been at the forefront of CAMD with quite a few established models. For instance, the preeminent OptCAMD framework tackles several complexity issues, and is capable of integrating machine learning methods for property prediction, demonstrating considerable success in several case studies [34], [214]. Further, a comparative study reported that genetic algorithms consistently perform as good as or better than present-day state-of-the-art deep learning models on molecular design tasks with much lower computational cost [205], [211]. However, rapid advances are expected to follow as the nascent deep learning in molecular design literature presents proof-of-concept demonstrations in inorganic solid-state functional materials and reticular frameworks [215], [216]. In the short term, we anticipate that hybrid systems involving decomposition, deep learning, and knowledge-based methods hold the most potential to solve problems of significance. In this context, many sources in the literature have addressed the efficacy of anchoring deep learning algorithms with scientific theory through a synergistic coupling of response and loss functions, selecting theory-compliant model architectures, constraining the space of probabilistic models with theory-based priors, and including theory-based regularization terms [29], [217]. Demonstrated successes of the fusion of deep learning and scientific knowledge include physics-informed neural networks [218], interfaces between quantum mechanics and neural networks [28], molecular deep tensor neural networks, SchNet [219], among others. As progress in the field of deep generative learning accelerates, we anticipate that more sophisticated methods will emerge for better integration of chemistry knowledge, resulting in improved performance and broad-ranging latent chemical spaces [220].

## 4.2. Property Data Availability

The issue of limited availability of property data plagues the development of property prediction models, which serve as the underpinning of the CAMD problem. The complexity of this issue is compounded when mixtures or products are considered. The collection of reliable property models can be enlarged using theory-based methods, data-driven methods, or their combinations [89]. Despite the groundbreaking leap that deep learning has brought to property prediction, this data-intensive class of methods poses data availability as a major limitation to establishing confidence in the generalizability of its predictive models [19], [20]. Further, although sizable datasets exist for several chemical properties, many central properties that are more expensive to compute or measure remain limited. A viable solution route is to generate pseudo-experimental property data using reference methods, such as quantum mechanics [221], DFT [20], and COMSO-RS [53], leading to larger model availability and data standardization [10], [89]. Also, different approaches could be explored to address the paucity of property data, including data fusion, transfer learning, active learning, and text mining [222], [223]. Future studies exploring the cost-accuracy tradeoff through the generation and testing of low and high-fidelity property data may offer insights into relating certain properties to elemental structures and their transferability to larger sets of molecular structures [224]. Given the successful applications of text and image-mining in extracting novel compounds and synthesis routes [138], [225], the task of extracting property data from published literature remains a potential route to substantially increase the volume of readily available data. Additional progress in property data mining and aggregation is expected to mitigate some of the challenges associated with the absence of engineering knowledge in structure-property relationships for different classes of molecules, such as crystal structures [226], alloys [227], proteins and nucleic acids [228], [229], polymers [230], ionic liquids [231], and biologics [232], [233]. This enlargement of open-source datasets and the library of available property models is vital to accelerating further development of this field [89].

## 4.3. Molecular Representations

As seen, vital to the success of the predictive and generative tasks is the choice of an expressive representation of molecules. There does not seem to be a single representation that fits all properties based on a rigorous study that controlled properties through varying elemental compositions and atomic configurations [234]. CAMD is often performed at multiple layers and different scales of complexity [31]. As distinctions between isomers or similar structures become more important, representations become more complex and mechanistic for improved properties prediction [9]. While 3D representations suffer from sparsity, invertibility, and invariances issues, there is a shift towards the use of 3D representations in the literature with several methods tackling such issues [156], [157]. Further, constructing chemical latent spaces from a combination of representations could lead to significant performance improvements. A heteroencoder encoding several representations was shown to outperform an encoder with a single representation input [143]. Thus, for molecular design frameworks with multiple design stages, it may be best to include several molecular representations, with each design stage assigned the most suitable representation for the given design task [31]. Further, the use of non-atom centric representation of molecules as 3D structure and electrons has been recently suggested to describe molecular systems more accurately [204]. Looking ahead, innovation with available representations and the development of novel ones appear to be promising creative endeavors.

## 4.4. Generative Models

The current pace and thrust of developments in generative deep learning present interesting new avenues for molecular generation and optimization. There is much to be accomplished from improving the generalizability of models and relaxing the i.i.d. assumption for reflecting real-world data, to covering unexplored areas of the chemical space, and performing operations in the space by transforming learned embeddings into practical rules. As the entire field of deep learning advances, novel methods applicable to our problem of interest emerge, such as multi-level VAEs [178], [235], Boltzmann Generator [236], and GraphNVP [237], [238]. Further, the literature points to many alternative routes and several extensions of interest. The use of token embeddings has been suggested to construct more robust RNN-based models. Another way to improve RNNs is through their modification to factor in more complex memory and attention mechanism [142], [239]. Even with the extensive success, the embeddings algorithm has achieved in natural language processing, it is also possible to adapt or tweak other algorithms as Transformer [240] and Sequence-to-sequence learning [241]. Provided that implementations with graphs as molecular representations assume graphs to be static, incorporating RNNs to build dynamic graphs was demonstrated to address their invertibility [180]. Also, two tree graph-based methods identify tree decomposition optimization as a route to improve their generalization using goodness criteria as the minimum description length [242] or Bayesian approaches [130], [173], [243]. With novelty as main scoring criteria, many implementations of the powerful framework of GANs still suffer from mode collapse. A possible approach to alleviate this issue is by incorporating recent developments in dynamically controlled attention RNN for text generation tasks [238], [244], [245]. Provided that the generator model in GANs is often trained as a stochastic policy in a reinforcement learning framework, several approaches that potentially lead to more stable training and better generators have yet to be applied to molecular design. These include, but are not limited to, maximum entropy inverse reinforcement learning [246], actor-critic methods [247], and multi-objective constrained reinforcement learning [248]. In light of the success of hierarchical representations of large molecules (polymers, proteins, and metal frameworks) [249], [250], hierarchical GANs may also have the capability to learn a domain-invariant feature representation of more complex and larger molecules [251].

## 4.5. Evaluation Metrics

The development of consistent tasks and general evaluation metrics to compare generative and predictive models and rapidly screen candidate molecules and products is essential to speed up material discovery. Adopted metrics should rigorously and independently assess the tasks of molecular generation and molecular optimization. Pioneering efforts in the development of evaluation metrics include the MOSES [204] and GuacaMol [205] platforms, which provide an open-source library with datasets, baseline models, and evaluation metrics. Yet, as many current metrics have been based on heuristics and rules of medicinal chemistry, novel benchmarks are required to holistically design molecular solutions for broader spectrums of applications and disciplines [204]. Further, with most proposed metrics predominantly focusing on the evaluation of generative models, it is imperative to develop metrics around the property prediction task, especially for quantifying uncertainty. This metric is particularly essential for opaque methods as neural networks in order to describe model ignorance and identify molecules for which experimental or pseudo-experimental data may be needed [252], [253]. As such, higher levels of collaborative work are needed to refine and extend the library of open-source benchmarking platforms and methods.

## 4.6. Integrated Product and Process Design

The application of deep learning to CAMD is merely a starting facet of a complex autonomous system that concurrently integrates the generation, optimization, and synthesis of chemical products. Even though many CAMD methods and tools have long been developed, only a limited number of models have been implemented in the industry. Additionally, there is no published work that concurrently considers synthesis routes and product/process design [89]. Deep learning models have been proposed for Computer-Aided Synthesis Planning (CASP) in order to accelerate the synthesis process. This type of models takes molecules as an input, and generates feasible synthesis routes with purchasable starting compounds [254]. Further exploitation of deep learning advances for binding the circle of design involves enabling autonomous experimentation and synthesis in self-driven laboratories [255]. As a pivotal step towards unleashing the "Moore's law for scientific discovery", the development of integrated design and synthesis systems requires synergy between theoretical and experimental researchers. [256].

The deployment of multi-scale modeling and efficient property estimation models within process design will lead to improved systematic methodologies for rapid prediction and evaluation in product/process design. As the field stands, the lack of systematic methods renders the evaluation process ineffective, excluding potentially superior molecules or mixtures [31]. In order to expand the portfolio of chemical products and improve the efficiency of processes, it is important to develop reliable, fast, and sustainable design and simulation tools. Hybrid product and process design systems linking knowledge-based decomposition methods with deep learning-based techniques would be necessary to approach current material design problems, such as polymers for membranes, zeolites for adsorbents, metals for catalysts, and enzymes for bio-catalysts. Such design problems, however, are less tenable due to the lack of property data and complexities in structure-property relationships. As seen, the role of researchers in the short run is to integrate learning-based models into the knowledge-based design of products and processes, offering a roadmap of knowledge gaps and challenges that needs to be addressed.

## 5. Conclusion

Although the knowledge-based CAMD methods have had longstanding success for systematic screening and identification of promising molecules, the emergence of deep learning approaches for molecular design holds a transformative potential in the near future. In this paper, we reviewed recent progress, limitations, and opportunities in CAMD for both knowledge-based and deep learning-based approaches. In addition to offering a detailed review of knowledge-based property prediction methods and solution techniques, the article also presented a survey of state-of-the-art deep generative CAMD models, examining various molecular representations, deep learning architectures, and evaluation benchmarks. The comparative descriptions of the key elements underlying the knowledge-based and data-driven CAMD methods revealed several challenges and opportunities from multiple facets. Building on the discussions of the current challenges, we identified a key promising path forward represented by hybrid methods, which harness the powerful capabilities of deep learning while leveraging the accumulated wealth of knowledge in the rich domain. Future work could be directed towards building large and diverse datasets of property data, developing more expressive molecular representations, advancing deep generative models unanchored in the current assumptions, establishing better benchmarking methods, and integrating products and processes in the design loop. As seen, success in these endeavors largely hinges on work and innovations around integrating the various forms in which relevant chemistry and physics theory and knowledge are represented.